# We the Droplets: A Constitutional Approach to Active and Self-Propelled Emulsions

Samuel Birrer[1#], Seong Ik Cheon[1#], and Lauren D. Zarzar[1,2,3]*
1. Department of Chemistry, The Pennsylvania State University, University Park, PA 16802 USA
2. Department of Materials Science and Engineering, The Pennsylvania State University, University Park, PA 16802 USA
3. Materials Research Institute, The Pennsylvania State University, University Park, PA 16802 USA

[#]These authors contributed equally
*correspondence to ldz4@psu.edu

# 1 Introduction

The mechanisms by which microscopic living and nonliving materials, such as bacteria and immune cells or active colloidal dispersions of particles and droplets, convert free energy into directional motion have been a keen interest of researchers for many decades.[1–4] Exploration of synthetic active materials has accelerated, inspired by biological self-propulsive swimmers and motivated by a desire to understand and design adaptive materials with behavior such as motility, autonomous response, collective interactions, cargo delivery, and emergent function.[3,5–11] Design strategies include creation of micro-objects with moving parts reminiscent of flagella or cilia,[12] as well as colloids that self-propel due to self-generated chemical gradients rather than mechanical actuation.[13] For instance, extensive research has been conducted on catalytically-powered nano and microparticles that exhibit self-diffusiophoresis or self-electrophoresis.[6,14] More recently, self-propelled liquid droplets and the collective interactions between active emulsion droplets has been of heightened interest.[4,15] Most self-propelled droplets are driven into motion by the Marangoni effect, wherein locally generated chemical gradients create interfacial tension variation along the drop surface. Understanding and controlling the active behaviors of droplets has proven challenging in many regards, given the complex coupling between chemical and hydrodynamic effects that necessarily arise. However, exciting recent advances in both theory and experiment within this interdisciplinary research area suggest that continued effort in understanding non-equilibrium dynamics of droplets will lead to impactful progress in understanding active matter and designing materials with life-like adaptive properties.

## 1.1 What is an active droplet?

Active droplets that spontaneously move in solution without the need for externally applied gradients are often called "swimming," "self-motile," or "self-propelled."[4,16] Many self-propelled droplets are spherical and compositionally isotropic, with exceptions such as Janus droplets. There are also droplets that are not self-propelled in isolation but become motile as a result of interactions with neighboring droplets in the emulsion.[17] Active emulsions interact with their surroundings through convective flows and chemical gradients, participate in molecular exchange at interfaces, and localize reagents and reactions, thereby allowing droplets[18] to engage in a different set of behaviors than active solid particles.[19,20] Droplets can exhibit versatile responsive or interactive behaviors such as clustering, assembly, chasing, stop-and-go motion, attraction, and repulsion.[21–23] Because of their liquid nature, dynamics, sensitivity to a range of chemical and physical stimuli, and primitive resemblance to biological matter, active microdroplets are of great interest for the development of responsive and autonomous materials with potential for complex behaviors such as emergent function, chemical communication, and self-preservation (**Figure 1**).[24]

## 1.2 Approach of the review

The field of active matter, and particularly active emulsions, is growing rapidly, with significant progress made recently on both theoretical/computational and experimental fronts.[4,15,19,25,26] In this review, we focus on summarizing the experimental efforts and progress related to active liquid droplets. We narrow our scope specifically to droplets that are fully immersed in a continuous fluid phase, paying special attention to the droplet chemical compositions and droplet structures. Wetted or sessile droplets which have a liquid-air or liquid-solid interface can also be active and motile,[27–29] but they require additional considerations and are not covered in this review. The constitution of active emulsions – the chemical composition of their phases and the structure of their interfaces – is deeply important in understanding and designing their properties and behavior. The chemistry and construction of droplets varies widely, including oil-in-water and water-in-oil drops, drops stabilized with ionic surfactants, nonionic surfactants, or particles, and drops containing a single phase or multiple phases with various internal morphologies. The interactions of dispersed liquid phases with one another, as well as the influence of surfactants and other additives, can govern the behavior of active droplets, including their mechanism of motion, speed, trajectory, chemical exchange between dispersed and continuous phase, and droplet lifetime. The diversity in chemical compositions and structures leads to an equally diverse range of properties and dynamics.

In this review, we begin by describing the chemical and structural composition of active emulsions, and we then relate composition to the resulting emulsion properties and mechanisms of propulsion. We consider not only traditional single emulsions but also more complex variants, such as Janus droplets, Pickering emulsions, and multiple emulsions. We consider the active behavior of isolated droplets, then discuss how collections of droplets interact with each other to generate pairwise or multibody collective behaviors. We highlight progress on understanding how active emulsions interact with physical barriers, walls, and confining environments that shape the local chemical gradients and fluid flow. Throughout the discussion, we highlight the chemical and structural elements of each experimental emulsion system and how these characteristics impact factors such as the propulsion mechanism, speed, droplet orientation, direction, and time-evolution of motion. Finally, we describe general research trends, offer insights on best practices for communicating experimental research results in this area, and provide perspective on the promising future research directions in the field.

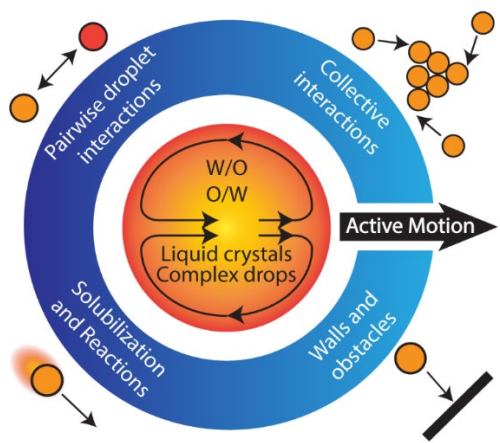

**Figure 1. Active emulsion droplets are diverse in composition and behavior.** Many different compositions of droplets have been shown to exhibit active behavior, including water-in-oil (W/O), oil-in-water (O/W), with isotropic phases and liquid crystal phases, and complex droplets including double emulsions and Janus drops. Active droplets can be self-propelled and/or interact with other droplets to give rise to pairwise interactions such as repulsion, attraction, and chasing. Collections of droplets can self-organize into clusters and patterns. Droplets can also respond to and interact with boundaries in the surroundings such as walls, pillars, or obstacles, which shape the local chemical gradients and fluid flow.

## 2 Chemical and Structural Elements of Active Droplets

### 2.1 Constitution of droplets

The constitution or composition of droplets begins with the chemistry of the liquid phases and their compatibility and spatial distribution.[30] An emulsion is a dispersion of one liquid within another and can be fabricated by mixing two or more immiscible fluid phases together, often in the presence of an interfacial stabilizer to prevent droplets from rapidly coalescing. Hence, most emulsions consist of a minimum of three materials: an inner or dispersed phase, an outer or continuous phase, and a stabilizer. The condition of phase immiscibility generally requires one phase to be more polar or hydrophilic (conventionally represented as W for water) and the other less polar or more hydrophobic (conventionally represented as O for oil). The two "traditional" types of emulsion droplets are thus oil-in-water (O/W) and water-in-oil (W/O). Rarer emulsion systems include water-in-water droplets (W/W), or oil-in-oil (O/O), and careful selection of the two liquids and stabilizer is needed to create emulsions of these types. Droplets of a single liquid phase dispersed with the continuous phase are called single emulsions. Other structures of droplets possessing more structural complexity also exist, which are often called complex droplets.[31] For example, Janus drops are complex droplets that contain two phases per droplet, where each phase shares an interface with the continuous phase; variations include biphasic Janus oil droplets in water (OO/W) as well as biphasic Janus aqueous droplets in oil (WW/O). Double emulsion droplets contain core-shell phases, such as water-in-oil-in-water (W/O/W), or oil-in-water-in-oil (O/W/O). Droplets can also vary widely in size. While most active droplet studies have been performed on microscale droplets (10-100 μm diameter) others focus on much larger droplets, on the order of one or even several millimeters in diameter.[32–34]

### 2.2 Droplet fabrication

Emulsions droplets are generated by mechanically shearing, mixing, or homogenizing immiscible liquids together. The means of emulsification and the stabilizers used largely determine droplet size and size dispersity. Bulk emulsification by shaking, vortexing, or sonicating the multiple phases and surfactant together is the easiest approach to making droplets, but the resulting drops are generally polydisperse in size. To fabricate monodisperse droplets of controllable size, microfluidics or membrane emulsification methods are used. Microfluidics is more common, is suitable for O/W and W/O, and enables the creation of complex droplets (double emulsions, Janus droplets) that otherwise would not be possible to fabricate uniformly.[35,36]

### 2.3 Surfactants and stabilization of emulsion droplets

Because the interfacial tension ($\gamma$) between most oils and aqueous phases is high, droplets are usually interfacially stabilized by the addition of surface-active agents (surfactants) or emulsifiers to prevent droplet coalescence and wetting.[37] In addition to small-molecule surfactants, polymers and nanoparticles are also commonly used to stabilize droplet interfaces. If particles are used, then the emulsions are known as Pickering emulsions.[38] Because surfactants directly influence interfacial tensions of droplets, and interfacial tension gradients are typically the driving force for drop motion, the specific type, chemistry, and concentration of surfactants used influences many aspects of the droplet structure, motility, and stability. Commonplace molecular surfactants that are used in the active droplet literature can be categorized as either ionic (e.g., sodium dodecyl sulfate, SDS) or nonionic (e.g., Triton X-100) and can be described with a Hydrophilic Lipophilic Balance (HLB) value. A higher HLB indicates a more hydrophilic surfactant and lower HLB indicates a more lipophilic surfactant. Typically, the surfactant should be more soluble in whichever liquid is intended to be the continuous phase. Hence, surfactants with higher HLB generally stabilize O/W emulsions more effectively (e.g., HLB=14.6 for Triton X-100), while surfactants with lower HLB stabilize W/O emulsions more effectively (e.g., HLB=4.3 for Span 80). Above a certain concentration, called the critical micelle concentration, or CMC, surfactants self-assemble into aggregates or micelles. Micelles can uptake the contents of droplets and transport them into the continuous phase, a process called solubilization.[39] The interfacial properties and dynamics of surfactants,[40] as well as the kinetics and mechanisms of micellar solubilization, are complex, and research is ongoing to elucidate how these interfacial processes affect droplet active behaviors.

## 2.4 Propulsion mechanisms of active droplets

A fundamental aspect of active droplets is the chemomechanical mechanism by which their motion and interactions are actuated. This review focuses on active droplets that can either self-propel or interact with other droplets via locally generated gradients without the need for externally applied fields. Hence, we will not be considering mechanisms of motion driven by application of magnetic fields, electric fields, thermal gradients, or spatially varying light intensity, although these stimuli can also be used to induce drop motion. Most self-propelling active droplets chemotax via the Marangoni effect, wherein sustained interfacial tension gradients across the droplet surface give rise to convective flows and motion. The Marangoni effect generates microscopic flow along the drop interface proceeding from low to high interfacial tension. Chemophoretic effects are also relevant[41,42] and become more influential when the droplet has high viscosity, although in most cases considered in the literature, Marangoni effects are expected to dominate.[43] Interfacial tension gradients across the droplet interface and Marangoni flows are commonly generated via two general approaches: chemical reactions of interfacially active chemical species (**Figure 2A**) and solubilization processes (**Figure 2B**).

## 2.5 General mechanism for chemical reaction-driven active droplets

The interfacial tension of the droplet interface depends on the chemical structures of the drop and continuous phase, the surfactant structure, and the interfacial surface coverage of the surfactant. Therefore, changes in the chemistry of the surfactant, such as via chemical reactions that can in-situ "improve" (increase the surface activity of) a surfactant or degrade (decrease the surface activity of) a surfactant, can be used to affect the interfacial tension. Spontaneous interfacial tension gradients that form along the droplet surface will induce motion via the Marangoni effect and feedback processes can sustain the droplet propulsion while the reaction proceeds (**Figure 2A,C**). In general, most chemical reactions that are used to drive droplet motion occur at or near the droplet interface. Reagents are often dispersed in both the continuous and dispersed phase such that they only have opportunity to react near, the drop interface. For example, in an active emulsion system of 4-octylaniline drops in water, the aqueous surfactant *N*-(4-[3-[trimethylammonio]ethoxy]benzylidene)-4-octylaniline bromide reacts with a catalyst dispersed in the oil drop and undergoes hydrolysis; the hydrolyzed surfactant is less effective, thus leading to interfacial tension gradients and Marangoni flow.[44] In some cases, where cascading reactions or reaction networks are used, there may be a reaction occurring within the dispersed or continuous phase that produces a secondary reaction at the drop interface.[45] For example, Suematsu et al.[45,46] demonstrated active water droplets in squalane that are fueled by the Belousov-Zhabotinsky (BZ) reaction which creates bromine that then reacts with a surfactant (monoolein) to improve the surfactants' interfacial activity and generate Marangoni flow. Controlling conditions such as the pH that may affect the kinetics of such reactions, as well as the surface activity of the products and reactants, can provide some control over droplet speed and lifetime.

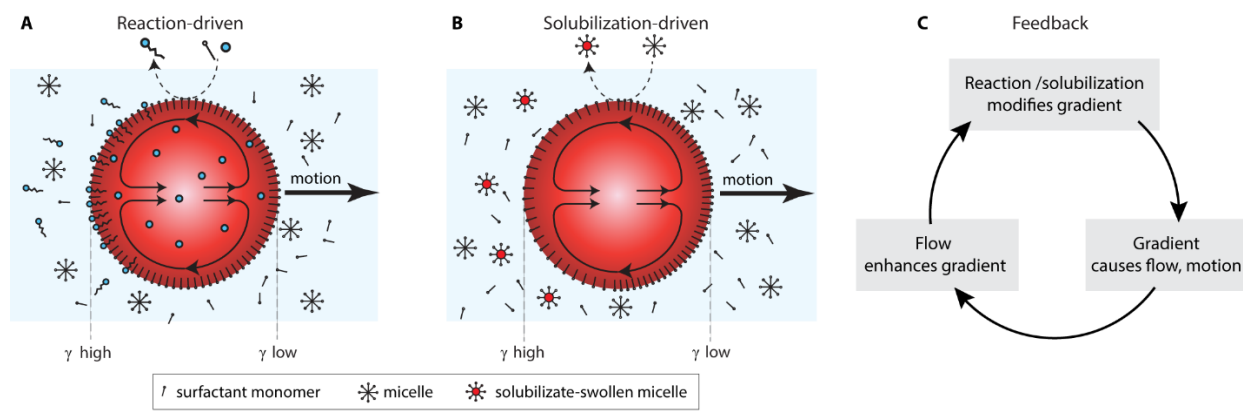

**Figure 2. Droplet motion is typically driven by interfacial tension gradients along the drop surface that generate Marangoni flow.** Interfacial tension gradients are usually created by either **A,** chemical reactions wherein an interfacially active molecule is created, modified, or degraded, or **B,** solubilization, an interfacial process wherein drop contents are transferred into micelles in the continuous phase. **C,** Feedback processes, wherein generated chemical gradients lead to flow, and flow in turn modifies the chemical gradients, enable an isotropic droplet to sustain self-propulsion. The schematic in **C** was inspired by [47].

### 2.6 General mechanism for solubilization-driven active droplets

Solubilization is a dynamic interfacial process by which droplet contents are transferred across the drop interface into the continuous micellar phase.[39] Depending on the surfactant molecular structure and the specific dispersed phase used, the kinetics of the solubilization process[48–51] as well as its effects on interfacial tension can vary. It is not yet fully understood how solubilization or the products of solubilization affect the interfacial tension, and the mechanisms may vary based on the emulsion composition (e.g. ionic vs nonionic surfactant, specific oils, etc.).[52] Some reports indicate that the presence of solubilizate may depress the CMC and that this is the source of interfacial tension gradients that drive drop motion.[53] Regardless of the specific mechanism, the solubilization process and/or gradients the solubilizate concentration clearly induce droplet motion via the Marangoni effect in many different emulsion compositions, which we highlight herein. Gradients in the distribution of surfactant solubilizate around the droplet, which are sometimes referred to as gradients of "empty" micelles and "solubilizate-swollen" or "solubilizate-filled" micelles, can be induced by mechanisms such as hydrodynamic instability, symmetry breaking from a substrate beneath the drop,[47,54,55] or by other nearby droplets, leading to sustained interfacial tension gradients and convective flow. For self-propelled droplets, higher concentrations of solubilizate in the micelle solution are correlated to higher interfacial tensions. As the droplet solubilizes, it also shrinks in size, which may also lead to changes in active behavior over the course of the droplet's lifetime.[17,56]

Despite the molecular complexity of the solubilization process, creating a self-propelling droplet that utilizes the solubilization mechanism is quite simple and at minimum requires appropriate choices of dispersed phase, continuous phase, and surfactant at sufficiently high concentration. Solubilization is effective for inducing motion in O/W and W/O single emulsion droplets and multiphase droplets. An example O/W system is diethyl phthalate (DEP) in water stabilized with SDS.[47] An example W/O system is water droplets in squalane stabilized with monoolein.[43] Given how experimentally straightforward it is to produce solubilizing active droplets, these emulsion systems have been an attractive target of study. Although sometimes not discussed in the literature, solubilization effects may also contribute to active droplet motion in cases where chemical reactions of the surfactant are also occurring, as described in the previous section.

**2.7 Other mechanisms for droplet propulsion**

Although most active droplets employ interfacial tension gradients to induce drop motion, there are examples of active droplets that self-propel by other mechanisms. In one example, phase transitions in the interface layer of an alkane oil droplet caused flagellar structures to extend reversibly from the surface, propelling the drop at speeds around 0.25-0.5 µm/s (one to two orders of magnitude slower than most active droplets propelling by solubilization or reactions).[57] Aqueous liquid crystal droplets containing living *B. subtilis* bacteria also demonstrated active motion in a nematic oil phase by a mechanism relying on turbulence generated by bacterial motility.[58] Movement was observed in actin-containing emulsions (≈0.6 µm/s), which are thought to generate Marangoni flows through the interaction of actin with the droplet interface.[59] The mechanism of motion was shown to be dependent on the nature of the substrate on which the droplets rest.

**Table 1:** Experimentally demonstrated active emulsion systems. Columns include the construction of the droplets (O/W, W/O, etc.), droplet composition, continuous phase composition (minus stabilizer), the stabilizer used, a designation of ionic (I) or nonionic (N) if the stabilizer is a surfactant, the general mechanism for propulsion, and approximate droplet diameter. Please refer to the Abbreviations list.

| Construction | Droplet Phase | Continuous Phase | Stabilization | Ionic or Nonionic | Mechanism | Approx. diameter (μm) | Ref |
|---|---|---|---|---|---|---|---|
| O/W | 1-decyne and HBA | Water and CuSO$_4$, hydrophilic azide | 4-(octyltriazole)benzoyl-oxyethelene-*N,N,N*-trimethylammonium bromide | I | Reaction (surfactant click chemistry) | 100 | 60 |
| O/W | HDA, spiropyran, toluene, DCE | Water and CaCl$_2$ | TTAB, SDBS, merocyanine | I | Reaction (UV photoisomerization) | 55 | 61 |
| O/W | 4-octylaniline | Water | hydrolyzable TAB surfactant | I | Reaction (surfactant hydrolysis) | 30 | 44 |
| O/W | 5CB and BPD | Water | TTAB | I | Solubilization | 30-100 | 62 |
| O/W | Aniline and ethanol | Water and salt | ethanol | N | Other (density gradient) | 2,000 | 63 |
| O/W | *N*-bromoalkanes, *n*-iodoalkanes | Water | Tergitols, Makon TD-12 | N | Solubilization | 10-200 | 52 |
| O/W | 1-bromooctane, EFB, 1-iodoalkanes | Water | Triton X-100, Capstone fluorosurfactant | N | Solubilization | 50 | 17 |
| O/W | DEP | Water | SDS | I | Solubilization | 15-45 | 47 |
| O/W | DEP | Water and NaCl | SDS | I | Solubilization | 30 | 53 |
| O/W | Dodecane, decane, others with M280 light absorbing particle | Water, PBS | Span 80 | N | Other (thermocapillary) | 2 | 64 |
| O/W | HBA | Water | CTAB, OTAB, HTAB | I | Solubilization | 103-232 | 65 |
| O/W | HBA | Water | DTAB, LOETAB-fatty acid | I | Reaction (surfactant hydrolysis) | 50-120 | 66 |
| O/W | Hexane, octane, nonane, decane, dodecane | Water and ammonia | SDS | I | Other (Marangoni effect) | 37-131 | 67 |
| O/W | nitrobenzene | Water | oleic acid | I | Reaction (surfactant hydrolysis) | 5,000 | 68 |
| O/W | nitrobenzene | Water and rare earth metal ions | DEHPA | I | Reaction (surfactant deprotonation) | 1,000 | 69 |
| O/W | nitrobenzene | Water and NaOH | DEHPA | I | Reaction (surfactant deprotonation) | 500 | 70 |
| O/W | nitrobenzene | Water | oleic acid | I | Reaction (surfactant hydrolysis) | 50 | 34 |
| O/W | octane | Water and ammonia | SDS | I | Other (Marangoni effect) | 70 | 71 |
| O/W | octane | Water and ammonia | SDS | I | Other (Marangoni effect) | 70-120 | 72 |
| O/W | octanol | Water | SDS | I | Reaction (surfactant reproduction) | 50 | 73 |
| O/W | Tetradecane, pentadecane | Water | Brij 58 | N | Other (flagellar) | 40 | 57 |
| O/W | Tributyrin and PDMS | Water | Protein (lipase) | - | Reaction (lipase consumption of oil) | 100 | 74 |
| LC/W | 5CB | Water | TTAB | I | Solubilization (rheotaxis) | 80 | 75 |
| LC/W | 5CB | Water | SDS | I | Solubilization | 50 | 76 |
| LC/W | 5CB | Water | TTAB | I | Solubilization | 30-45 | 77 |
| LC/W | 5CB | Water | TTAB | I | Solubilization | 20-200 | 78 |
| LC/W | 5CB | Water and glycerol, other solutes | TTAB | I | Solubilization | 20-100 | 56 |
| LC/W | 5CB | Water and $D_2O$ | 4-Butyl-4'-(6-trimethylaminohexyloxy) azobenzene, PVA | N | Solubilization (light-tuned) | 35-185 | 79 |
| LC/W | 5CB | Water and $D_2O$ | TTAB | I | Solubilization | 45 | 21 |
| LC/W | 5CB + BPD (10:1) | Water | TTAB | I | Solubilization | 50 | 80 |
| LC/W | 5CB and BMAB/BHAB | Water | SDS | I | Solubilization | 100 | 81 |
| LC/W | 5CB doped w/ CB15, Me-m, Ph-m | Water | TTAB | I | Solubilization | 10-23 | 82 |
| LC/W | 5CB doped w/ R811 | Water | TTAB | I | Solubilization | 15 | 83 |
| LC/W | 5CB, 5CB and BPD, CB15 | Water | TTAB | I | Solubilization | 15-50 | 84 |
| LC/W | 5CB, 8CB, MBBA | Water | TTAB | I | Solubilization | 100-200 | 85 |
| LC/W | CB15 | Water | TTAB | I | Solubilization | 50 | 22 |
| LC/W | CB15 | Water | TTAB | I | Solubilization | 50 | 86 |
| LC/W | CB15 | Water | TTAB | I | Solubilization | 50 | 87 |
| LC/W | CB15 | Water and $D_2O$ | TTAB | I | Solubilization | 48 | 88 |
| LC/W | CB15 | Water and glycerol | TTAB | I | Solubilization | 60 | 89 |
| LC/W | MBBA | Water | SDS | I | Solubilization (UV-induced) | 50 | 90 |
| LC/LC | 5CB | Water and DSCG | SDS | I | Solubilization | 25 | 91 |
| W/O | HEPES buffer and actin | HFE-7500 | PEG fluorosurfactant, Krytox | N | Other (Marangoni-driven rotation) | 40 | 59 |
| W/O | Water | Squalane | monoolein | N | Reaction and solubilization | 550-1,000 | 46 |
| W/O | Water | Squalane | monoolein | N | Reaction (monoolein bromination, BZ) | 30-600 | 92 |
| W/O | Water | Squalane | monoolein | N | Reaction (monoolein bromination) | 1,000 | 45 |
| W/O | Water | Squalane | monoolein | N | Solubilization | 100 | 93 |
| W/O | Water | Squalane | monoolein | N | Solubilization | 30-400 | 94 |
| W/O | Water and $NaBrO_3$ and $H_2SO_4$ and ferroin and others | Oleic acid | oleic acid | I | Reaction (BZ) | 1,000 | 95 |
| W/O | Water and $NaBrO_3$ and $H_2SO_4$ and ferroin and others | Oleic acid | oleic acid | I | Reaction (BZ) | 1,000 | 96 |
| W/O | Water and $NaBrO_3$ and $H_2SO_4$ and ferroin | Squalene | monoolein | N | Reaction (monoolein bromination, BZ) | 1,000- 5,000 | 33 |
| W/O | Water, water and salt | Squalane, tetradecane | monoolein | N | Solubilization | 35-60 | 43 |
| W/O | Water and $NaBrO_3$ and $H_2SO_4$ and ferroin | Squalane | monoolein | N | Reaction (monoolein bromination) | 1-2 | 97 |
| LC/O | Water and kinesin and tubulin | HFE-7500 | PFPE-PEG-PFPE | N | Other (active nematic) | 30-60 | 98 |
| LC/LC | Water and kinesin and tubulin and PEG | 5CB | Tween 80, PF-127 | N | Other (active nematic) | 100 | 99 |
| LC/LC | TB, TB and DSCG and *B. subtilis* | MAT-03-382 | egg yolk lecithin | - | Other (LC + bacteria) | 10-140 | 58 |
| LC/LC | TB and LCLC and DSCG and *B. subtilis* | 5CB | lecithin | - | Other (LC + bacteria) | 60-100 | 100 |
| W/LC/W | 5CB (water) | Water | TTAB | I | Solubilization | 75 | 101 |
| O/W, W/O/W | Nitrobenzene and HDA and Oil red dye | Water | C12-HPTS, DDAB, HDA | N | Reaction (photochemical with light gradient) | 3,000 | 102 |
| O/W | Bromodecane | Water | Particles (silica) and Triton X-100 | N | Solubilization | 20-100 | 55 |
| W/W | Dextran-rich water | PEG-rich water | Particles (liposomes) | - | Other (polymer concentration gradient) | 5-120 | 103 |
| OO/W | Haloalkanes, EFB, fluorosilicone | Water | Triton X-100, Capstone fluorosurfactant | N | Solubilization | 50 | 23 |
| O-LC/W | 5CB-silicone oil | Water | SDS, PVA | I, N | Solubilization | 10-100 | 104 |
| O-LC/W | Perfluorobenzene-E7 nematogen | Water | SDS, TTAB | I | Solubilization | 50 | 105 |
| WW/O | Water-ethanol | Squalane | monoolein | N | Solubilization | 150 | 106 |
| WW/O | Water-ethanol | Squalene | monoolein | N | Other (Marangoni effect) | 30-80 | 107 |
| WW/O | Water-ethanol and DNA | Squalane | monoolein | N | Solubilization | 35-85 | 108 |

# 3 Recent Experimental Progress in Active Droplets

Given how important the chemical constitution and structure of droplets are to the droplet motility, we aimed to compile a comprehensive list of experimental conditions and emulsion compositions that have been reported. **Table 1** organizes recent literature on active droplets by the droplet structure (O/W, W/O, W/O/W, etc.), surfactant identity and whether it is ionic or nonionic, droplet phase chemistry, propulsion mechanism, and approximate droplet size. By collecting this information, we hope to reveal trends in the experimental efforts related to active droplets, provide guidance on how the chemical composition of the droplets and structure of droplets influence active behaviors, and shed light on some underexplored chemical spaces that would be worth future consideration. Here, we discuss in more depth the literature compiled in **Table 1**, considering each class of emulsion structure.

## 3.1 Water-in-oil active droplets

*Isotropic aqueous droplets in oil.* W/O droplets make up a minority of active droplets studied and have less diversity in chemical composition compared to O/W. Aqueous droplets are usually dispersed in a long-chain hydrocarbon such as squalane, squalene, or tetradecane (**Table 1**). Sometimes cosolvents or solutes are added in the aqueous phase, such as ethanol or salts. W/O emulsions are commonly stabilized using oil-soluble surfactants, such as monoolein or Tween 80. Active behavior can be generated by solubilization (**Figure 3A**) and/or reactions that occur in either the continuous or dispersed phase (**Figure 3B,C**). Water droplets in 25 mM of monoolein surfactant of about 20 – 60 μm diameter undergoing solubilization move at 10 – 50 μm/s.[43] Reaction-driven active water droplets in squalane[45] can self-propel via the bromination of monoolein surfactant; sodium bromate was dissolved in the aqueous droplet and reacted with sulfuric acid (1.0 – 1.8 M) to form bromine which interfacially-reacted with monoolein dispersed in squalane forming a surfactant that generated lower interfacial tension. For a 1 mm droplet, without any acid, the speed was around 1 μm/s but increased to 5 – 30 μm/s with increasing acid concentrations within the droplet. A similar approach of monoolein bromination has been used to couple the motion of W/O droplets the oscillating BZ reaction (**Figure 3B,C**).[33,46,92]

*Liquid crystals in isotropic oil.* There are few examples of active emulsions with an oil continuous phase and liquid crystalline dispersed phase. Those reported consist of complex bioinspired mixtures of the proteins kinesin and tubulin, forming a liquid crystalline aqueous dispersed phase in either isotropic or liquid crystalline oil.[98,99] Kinesin, a motor protein, binds to tubules assembled from tubulin, forming an active gel that is fueled by ATP. The activated gel can generate internal flows within droplets, resulting in self-propelling nematic defects in the droplets,[99] as well as slow self-propulsion of the droplets themselves in some cases (typically <1 μm/s).[98] It is important to note that in these systems the term "active" is commonly used to describe multiple levels of motility: the ATP-powered walking of kinesin, the active flow in the gel-containing nematic phase, and the translational activity of droplets.

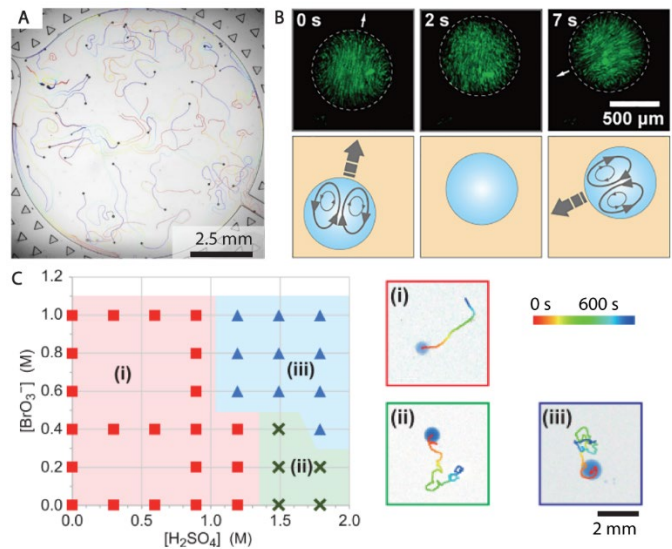

**Figure 3. Water-in-oil active droplets driven by solubilization and reactions**. **A**, Optical micrograph of solubilizing water-in-squalane droplets stabilized by monoolein with the droplet trajectories traced over 500 seconds. Reproduced from [43], copyright American Physical Society. **B**, Fluorescence micrographs of aqueous droplets in oil containing fluorescent particles allow visualization of the streamlines associated with Marangoni flow inside each drop. A corresponding schematic correlates how these streamlines are associated with droplet motion. These droplets are made active through a reaction-driven mechanism. Reproduced from [46], copyright American Chemical Society. **C**, Three different general trajectories of O/W droplet motion were observed when using different concentrations of reagents in a system of Belousov-Zhabotinsky reaction-driven emulsion. Key: (i) ballistic, (ii) fluctuating ballistic, (iii) random motion. Reproduced from [46], copyright American Chemical Society.

**3.2 Oil-in-water active droplets**

A majority of the active emulsion literature, as compiled in **Table 1,** has focused on O/W droplets. Within the O/W drop literature, two general categories of oils have been investigated: those that are isotropic, like alkanes, haloalkanes, and aromatic oils, and those that are liquid crystalline (**Figure 4**). Liquid crystals are distinguished from isotropic liquids in that they have long-range molecular orientational order, which can imbue active droplets with distinct properties.[109] We consider each of these classes of droplets separately.

*O/W emulsions with isotropic oils.* Aromatic oils (e.g., nitrobenzene, diethylphthalate, heptyloxybenzaldehyde) and nonaromatic oils (e.g., haloalkanes such as bromooctane or iodododecane, and alcohols) have featured largely in active droplet studies. In some cases, nematogens are mixed with isotropic oils to create isotropic oil-in-water emulsions; for example, 4-cyano-4'-pentylbiphenyl (5CB) when mixed with bromopentadecane (BPD) which eliminates effects associated with the nematic phase.[62] Notably, few reports[17] use alkanes. The choice of oils, such as the use of haloalkanes instead of alkanes, may be due to density considerations; alkanes are less dense than water thus requiring a slightly more complex experimental setup with an enclosed sample chamber such that droplets do not encounter an air-water interface. We find that ionic surfactants have been more commonly used than nonionic surfactants to stabilize isotropic oils that are active under solubilization conditions, although both are suitable. Example ionic surfactants include sodium dodecyl sulfate (SDS) (an anionic surfactant) and various n-alkyl trimethylammonium bromide (TAB) surfactants (all cationic surfactants): hexyl (HTAB), octyl (OTAB), dodecyl (DTAB), tetradecyl (TTAB), cetyl or hexadecyl (CTAB).[65,66] Nonionic surfactants that have been used to generate active behavior via solubilization include Triton X-100,[17,23,55] di-(2-

ethylhexyl)phosphoric acid (DEHPA),[69,70] Capstone FS-30 (a fluorosurfactant),[17,23] and Tergitol NP-9.[52] The surfactants are all used above their CMC, often at concentrations orders of magnitude above the CMC and sometimes as high as >20 wt%.

The specific choice of isotropic oil and surfactant for a solubilization-driven propulsion mechanism is important. Solubilization rates are greatly influenced by choice of oil, surfactant, and surfactant concentration. The same factors also impact droplet lifetimes, which range from minutes to hours depending on initial droplet size and solubilization kinetics. Some large hydrophobic oils, like iodohexadecane, do not solubilize at appreciable rates even at surfactant concentrations orders of magnitude above the CMC, and hence often do not generate sufficient solubilizate gradients to sustain active behavior. Exceptions exist, such as when the iodohexadecane is anisotropically modified by the addition of adsorbed surface particles.[55] Some surfactants are effective at stabilizing O/W emulsions but not effective at oil solubilization, such as nonionic nonylphenol ethoxylate surfactants with large ethylene oxide headgroups.[52] Hence, these surfactants are also not as useful to generate O/W active droplets.

O/W emulsions can also be driven by reactions, either in conjunction with or instead of solubilization. In some cases, the reaction increases the surface activity of one component of the system, thereby reducing interfacial tension locally where the reaction occurs. Methods used to achieve surfactant improvement include: hydrolysis of oleic anhydride to oleic acid,[34,68] hydrolysis of lauroyloxyethylene-*N,N,N*-trimethylammonium bromide to a catanionic surfactant complex,[66] deprotonation of di-(2-ethylhexyl)phosphoric acid to a phosphate,[69,70] and an interfacial click reaction of a head and tail to form the surfactant 4-(octyltriazole)benzoyl-oxyethelene-*N,N,N*-trimethylammonium bromide.[60] Light-responsiveness has also been incorporated using the photoisomerization of spiropyran into the slightly surface-active compound merocyanine, though its activity is not great enough to stabilize droplets alone.[61] In addition to improvement, surfactant deterioration can also give rise to motion, as evidenced by the hydrolysis of a "fuel" surfactant, *N*-(4-[3-[trimethylammonio]ethoxy]benzylidene)-4-octylaniline bromide into two less surface-active species.[44] Alternatively, some elements of reaction-based motility may be achieved without surface tension gradients, as in the case of oil droplets containing lipase, which hydrolyzes the triglyceride tributyrin into fatty acids with lower density, generating buoyant force to propel the droplet upward.[74]

*O/W emulsions with liquid crystals.* Although liquid crystalline character is not a prerequisite for active behavior, liquid crystals have been widely used in active droplet research. Liquid crystalline phases, particularly nematic phases in either the dispersed or continuous phase, are of interest because the long-range molecular ordering can sometimes give rise to different modes of motion than occur in isotropic liquids. Among the most common LC oils used are biphenyl-based compounds 4-cyano-4'-pentylbiphenyl (also called pentylcyanobiphenyl or 5CB), 4-*N*-octylcyanobiphenyl (8CB), 4-cyano-4'-(2-methylbutyl)-biphenyl (CB15), and *N*-(4-methoxybenzylidene)-4-butylaniline (MBBA), which form nematic phases.[85] Unlike optical imaging of isotropic oil drops, liquid crystal droplet internal molecular order and orientation can be visualized with polarized light microscopy. Changes in the liquid crystal structure as a function of modifications to the surfactant phase and droplet activity or fluid flow can thus often be visualized directly which can be advantageous for interpreting the flow fields of moving droplets (**Figure 4A,B**). The surfactant most commonly used to stabilize active liquid crystalline droplets in aqueous solutions is tetradecyltrimethylammonium bromide (TTAB), which is often used in concentrations between 7.5 and 25 wt%, far above the CMC of about 0.13 wt% (**Figure 4C**).[77] A high concentration of TTAB (>5 wt%) is required to generate any active motion in 5CB,[85] with droplet speeds of around 10 μm/s at 7.5 wt% and speeds plateauing near 20 μm/s at 15-25 wt%.[25] SDS has been used to stabilize active liquid crystal emulsions in similarly high concentrations of 10-25 wt%.[76] We find that other surfactants, especially nonionic surfactants, are rarely reported to be used at such high concentrations, although perhaps they have not been tested. Active liquid crystal droplets stabilized by nonionic surfactants have not been reported in the literature to the best of our knowledge, but we expect such a combination to be possible;

what is unclear is whether the liquid crystalline nature of the oil would have any impact on the properties of the droplet beyond what has already been observed for isotropic oil droplets. Viscosity modifiers such as glycerol have been added to the continuous phase to tune the Péclet number, which characterizes the ratio between advective and diffusive transport, and alter the directionality of the liquid crystal droplet trajectory.[89] However, follow-up work indicated that glycerol induces chemical changes beyond simply changing the viscosity, and use of viscosity-modifying polymers like polyacrylamide yielded different active droplet results;[56] glycerol, for example, is surface active and alters the surfactant micelle structure which can affect solubilization kinetics.[110] As such, care must be taken in interpreting the results of experiments when new chemicals are added. It is difficult to change one physical variable, such as density or viscosity, independently of the chemistry.

The liquid crystalline character of the oil is not central to the generation of active motion, which is usually caused by oil solubilization independently of the anisotropy of the dispersed phase. However, there are cases in which the liquid crystal properties influence the mode or pattern of droplet propulsion.[85] Added chiral compounds such as CB15 or isosorbide hexyloxybenzoate (R811) can also modulate active motion by creating handed defects in the nematic ordering of the dispersed phase. For example, there are liquid crystal droplets which rotate in spirals with a handedness that is tunable by choice of chiral dopant (**Figure 4D**).[82,83] Chiral liquid crystal is not needed to induce spiraling, which has also been observed in 5CB droplets, although in such a case there is no preference for spiral handedness.[77]

In addition to a liquid crystal dispersed phase, it is also of value to consider the motility of droplets in a liquid crystalline continuous phase. A unique liquid crystal-in-water active emulsion was made by Nayani, Córdova-Figueroa and Abbott by dispersing liquid crystalline 5CB droplets in a liquid crystalline aqueous phase of 17 wt% disodium chromoglycate.[91] They found that while isotropic oils prefer to propel along the expected direction of lowest resistance (that is, low effective viscosity from molecular orientation), nematic droplets prefer a direction of high resistance, because of the way their radial symmetry breaks and orients the droplet.

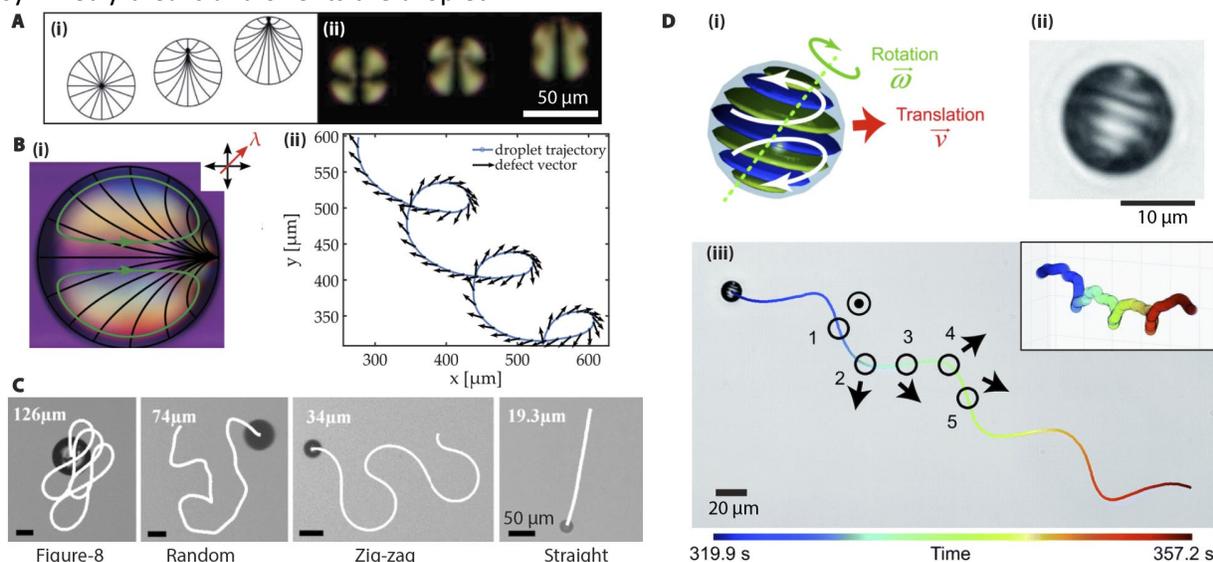

**Figure 4. Liquid crystal active droplets. A(i)**, Schematics and **A(ii)** polarized light micrographs of differently oriented nematic 5CB droplets having a point defect. Reproduced with permission from [25] under a Creative Commons Attribution 3.0 Unported License. **B(i)**, The distorted nematic director field (black lines) and Marangoni flow (green arrow lines) overlaid with a polarized light micrograph of a 50 μm diameter 5CB drop. The orientation of crossed polarizers and the λ retardation plate used for imaging are shown with the arrows in the top right. Reproduced with permission from [77], copyright American Physical Society. **B(ii)**, The curling trajectories of a 50-μm diameter nematic 5CB LC droplet solubilizing in TTAB

surfactant. Reproduced with permission from [77] copyright American Physical Society. **C,** Four modes of motion arising in 5CB droplets of varying size in 10 wt% TTAB. Reproduced with permission from [78], copyright American Physical Society. **D, (i)** Chiral cholesteric liquid crystal droplet schematic, **(ii)** droplet optical micrograph, and **(iii)** optical micrograph time-sequence trace of drop motion while the chiral liquid crystal is undergoing solubilization in TTAB surfactant. The droplet moves in a helical trajectory which is visualized in 2D and 3D (inset). (The numbers in **(iii)** correspond to labeled droplet images which are not reproduced here.) Reproduced with permission from [83], copyright Royal Society of Chemistry.

### 3.3 Complex and composite active droplets

Complex multiphase drops[111] and composite droplets, such as those containing particle additives, are of increasing interest due to the ability to use droplet structure, shape, and anisotropy beyond a simple spherical shape to modify droplet motility. The greater structural and chemical diversity of complex droplets can also lead to unique collective behavior and stimuli-responsive properties[111] that are not accessible in single phase droplets.

*Janus droplets.* Janus droplets are characterized by having two drop-internal phases that each have an interface with the continuous phase (**Figure 5A**). Janus droplets can be made with either an oil or aqueous continuous phase. The morphology of a Janus droplet is determined by the balance of the three different interfacial tensions at the droplet surface,[112] and the Janus droplet shape affects the droplet propulsion.[23,105] The propulsion mechanisms of Janus droplets are similar to those of single-phase emulsions in that they can still be driven by mechanisms of solubilization; in principle, Janus droplets can also be fueled by reactions but to the best of our knowledge, there are no examples reported. Because there are two phases inside the Janus droplet, the chemical pairing of the two drop-internal phases, as well as the compatibility with the surfactant, are important to consider. For solubilization-driven Janus drops, chemical partitioning of the solubilizing compound between the two phases of the Janus droplet plays a significant role in governing drop speed as well as direction of motion.[23] The mechanisms behind solubilizing Janus droplet propulsion are also closely related to those of "chasing" interactions between single emulsion droplets;[17] such binary droplet interactions are discussed in more detail in Section 4.1.

Active OO/W Janus droplets consist of two immiscible hydrophobic oil phases dispersed in an aqueous phase. There are many reported examples of such active Janus drops, some of which contain two isotropic oils, and some which contain both a liquid crystal and isotropic oil. In one embodiment, microscale droplets were composed of a perfluorobenzene-rich phase combined with E7 nematic liquid crystal-rich phase (which contains a mix of 5CB, 7CB, 8OCB, and 5CT) stabilized by TTAB.[105] These droplets were initially single-phase and isotropic, but over time, preferential solubilization and removal of select components from the droplet caused drop-internal phase separation to create a Janus drop with nematic regions. Over the course of the drop lifetime, changes in the drop morphology led to different dynamic states of droplet motion (**Figure 5Ai**). Another hybrid isotropic and liquid crystal droplet contained silicone oil combined with 5CB.[104] These droplets were stabilized with SDS, TTAB and PVA and move at about 10 μm/s with the 5CB phase facing forward. Over a dozen combinations of isotropic oil-containing Janus drops stabilized in Triton X-100 have been demonstrated.[23] Most contain a brominated or iodinated alkane (which solubilizes) paired with a fluorinated oil (which does not solubilize). Depending on the specific oil combination, such droplets can move several hundreds of microns per second in 0.5 wt% Triton X-100 and they typically propel directionally with the haloalkane in the rear, fluorinated oil in the front (**Figure 5Aii**). The partitioning of the haloalkane across the drop-internal oil-oil interface was shown to be a key factor governing the drop speed and direction; drops reversed their swimming direction when the partitioning of the haloalkane into the fluorinated oil was very low.[23]

Active WW/O Janus droplets consist of two immiscible polar or aqueous-rich phases dispersed in an oil phase. There are fewer examples of such droplets. One system of this type involves water and ethanol as the dispersed phases, with squalane or squalene as the continuous phase.[108] Uptake of the surfactant monoolein into the ethanol/water phase caused demixing of the droplet phases, which are ordinarily miscible. This type of Janus emulsion system has been studied as a vehicle for precipitation and extraction of DNA.[107]

*Double emulsions.* Multiple emulsions involve a layered structure with droplets nested inside one another. W/LC/W active emulsions (**Figure 5B**) were observed to have a "shark-fin meandering" trajectory that was influenced by the motion of the inner aqueous compartment as the entire drop self-propels.[101] Yucknovsky et al. have also prepared active double emulsions, consisting of a nitrobenzene oil droplet containing an aqueous droplet with 8-hydroxypyrene-1,3,6-trisulfonic acid (HPTS), which made the emulsion light-responsive.[102]

*Particle-containing active droplets.* Active droplets can be combined with solid particles to create a composite solid-liquid architecture. Solid particles add a new source of asymmetry and route to structure multiphase[55] active droplets and can significantly influence the speeds and droplet behaviors. Pickering emulsions, wherein droplets are stabilized solely by particles, are extensively studied.[38,113] However, active droplets containing particles differ in that these emulsions may also contain molecular surfactant, which facilitates droplet propulsion surfactant-mediated solubilization.[55] Cheon et al. investigated the self-propulsion of bromoalkane oil droplets with varying surface coverages of silica particles a range of concentrations and types of aqueous surfactants (Triton X-100, SDS, CTAB) with and without added salt.[55] The authors discovered that the addition of interfacially adsorbed particles led, in some cases, to order-of-magnitude increases in drop speed depending upon the oil, surfactant, and particle surface coverage. For example, without particles, bromodecane droplets tens of microns in diameter self-propelled at 50 µm/s in 0.5 wt% Triton X-100, but speeds increased to 250 µm/s when adsorbed particles covered approximately half the drop surface area. Particles packed at the rear pole of the drop on the oil–water interface, and as the drop solubilized, the particle surface coverage increased which modified the drop speed (**Figure 5C**). The mechanism by which the particles contribute to the enhanced speeds is still not fully understood, but it is a useful strategy to impart motility to droplets which would be otherwise inactive.

Another example of a particle-stabilized active drop is that of a W/W droplet, which Zhang et al. designed using an aqueous two-phase system (ATPS); one phase was rich in the polymer polyethylene glycol (PEG), the other phase rich in dextran, and the droplet was stabilized by liposome nanoparticles.[103] These droplets, however, did not spontaneously propel but rather chemotaxed in response to an applied concentration gradient of PEG and dextran in the continuous phase. The liposomes distribute to the back of the droplet and were ejected over time as the droplet chemotaxed. Yet another way in which solid particles can be used for active emulsions is to couple a droplet to a solid particle of a similar size. One such swimmer was fabricated with a light-absorbing particle adjoined to a droplet of n-dodecane, each on the order of 5 µm in diameter.[64] These hybrid droplet-particle swimmers were propelled in the presence of light via a mechanism of thermocapillary action, while the liquid compartment served to provide asymmetry to the swimmer shape.

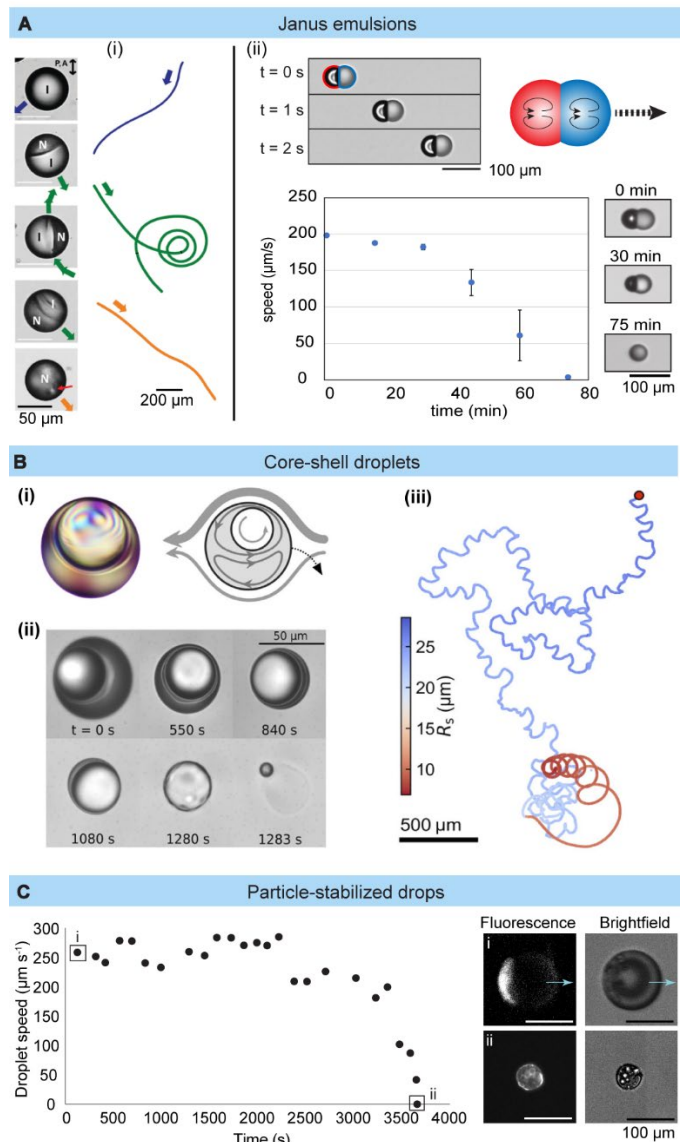

**Figure 5. The structure of complex droplets affects motion. A, (i)** Perfluorobenzene and E7 liquid crystal droplets in aqueous TTAB transform into different complex droplet geometries as the drop composition changes over time due to selective oil solubilization. The various morphologies of drops, which combine coexisting nematic (N) and isotropic (I) phases, exhibit different trajectories of motion as given by the colored trails. The red arrow in the bottommost image points to the defect in the nematic drop. Reproduced with permission from reference [105], copyright Royal Society of Chemistry. **(ii)** Janus droplets containing isotropic iodododecane (red) and ethoxynonafluorobutane (blue) self-propel in 0.5 wt% Triton X-100. Over time, the iodododecane lobe of the droplet solubilizes and shrinks, leading to a decrease in droplets speed. Reproduced with permission from reference [23], copyright Elsevier Inc. **B, (i)** Polarized micrograph of a water-in-5CB-in-water core-shell droplet stabilized by TTAB (left) and a schematic showing the flow within the core and shell of the droplet as it moves. **(ii)** Optical micrographs of the same type of droplet as in **(i)** but over time as the shell phase solubilizes. Eventually, the droplet shell bursts leaving behind a single emulsion droplet. **(iii)** The trajectory of a core-shell droplet over a 40-minute period. The transition from blue to red corresponds to the shell bursting. Reproduced with permission from [101] under the Creative Commons Attribution 4.0 International license. **C**, Adsorbing silica

nanoparticles to the surface of solubilizing oil-in-water droplets increases droplet speed. Here, fluorescently labeled silica particles were adsorbed to bromodecane droplets that were additionally stabilized by 0.5 wt% Triton X-100. Over time as the bromodecane solubilizes, the particle surface coverage increases, and the droplet speed slows. Reproduced with permission from [55], copyright Royal Society of Chemistry.

# 4 Droplets and their surroundings

For a droplet to chemotax, there must be asymmetry in the chemical surroundings. As mentioned previously, even isotropic, spherical droplets can spontaneously move without externally applied fields via mechanisms that locally induce chemical gradients, such as hydrodynamic instabilities, which in turn produce interfacial tension gradients and give rise to sustained motion. Chemical and physical asymmetry is inherent within many experiments and systems, even unintentionally. For example, microscale droplets will sink or float unless density matched with the continuous phase, and hence droplets are often studied when localized near a substrate or interface. The presence of the substrate or interface modifies the chemical gradients around the droplet as well as the hydrodynamics and provides a source of asymmetry that can affect convective flows. As such, even some droplets that may not be self-propelled laterally can still "pump" fluid in a directional manner[52,55] and the drop may levitate above the substrate.[47] Similarly, droplets can also interact with walls or pillars in their surroundings. Emulsions are also often studied as collections of droplets rather than in isolation, and the chemical gradients produced by droplet neighbors, even those many body lengths away, can lead to longer range interactions and directional motion even for droplets that are not self-propelled when in isolation. Chemical "trails" left behind by self-propelling droplets also influence the motion of other active drops that come into the vicinity of the trails. Changes in surroundings can be exploited to modify the active behaviors of emulsions in specific ways and to probe how collective droplet behaviors evolve with these external perturbations. We find increasing reports exploiting these collective and environmental effects in the literature, and we suspect this will be a significant focus of future research endeavors.

### 4.1 Pairwise droplet interactions

Depending on the mechanism of drop motion (e.g., solubilization or reaction-driven) the ways in which the droplets interact with each other can also differ. For active droplets reliant on chemical reactions, and particularly when one of the reactants is dissolved in the continuous phase, droplets will typically move away from other droplets that are competing for the same reactants[44] unless the reaction produces a product that lowers interfacial tension. For example, as described in Toyota et al., interactions between oil droplets (4-octylaniline) that consume surfactant *N*-(4-[3-[trimethylammonio]ethoxy]benzylidene)-4-octylaniline bromide as fuel have been shown to move away from each other in a pairwise interaction.[44] Kumar et al. demonstrated enhanced water droplet propulsion speed and a mix of attractive and repulsive drop interactions by leveraging bromination of monoolein with Belousov-Zhabotinsky (BZ) oscillatory reactions within the water.[33] The concentration of bromine in the BZ reaction oscillates over time, and droplets can be either attractive or repulsive depending on the timing of the oscillatory reaction and the distance between them.

Droplets that propel via solubilization can also interact in a pairwise fashion, and the interactions depend largely on the chemistry of the droplets. Typically, solubilizing self-propelled droplets are motile because they are "repelled" by their own solubilizate and therefore move towards regions of lower solubilizate concentrations and higher concentrations of "empty" micelles. So, if there are two solubilizing droplets of identical composition in the vicinity of each other, they repel (**Figure 6A**). This repulsion was demonstrated and quantitatively analyzed between two diethyl phthalate droplets stabilized by SDS that were initially forced into proximity with an optical trap.[47] Upon release of the trap, droplets were found to repel one another with a force that scaled with inter-droplet distance, $F \sim 1/r^2$. It has also been reported that droplets that are not self-propelled but still solubilizing can attract one another in cases where solubilized oil leads to reduced interfacial tensions.[52] The mechanisms by which solubilization raises or lowers interfacial tension is still being investigated.

If pairwise interactions occur between droplets of different composition, then non-reciprocal interactions can occur. Meredith and Moerman et al. described chasing interactions between microscale bromooctane drops (which solubilize) and ethoxynonafluorobutane oil drops (which do not solubilize) in 0.5 wt% Triton X-100 (**Figure 6B**).[17] When in isolation, the droplets slowly drifted at <3 μm/s and were not considered self-propelled. When the two different types of droplets came within several body lengths of each other, the bromooctane "predator" accelerated towards the ethoxynonafluorobutane "prey", eventually forming a touching pair that moved at >20 μm/s. Many other combinations of iodoalkane and alkane oils exhibited the same chasing behavior. This striking phenomenon is due to non-reciprocal oil exchange and the chemical ripening process occurring between a solubilizing "source" drop and a "sink" drop that uptakes the solubilized oil from the source. This work highlights the fact that even droplets with minimal activity alone, in more diverse chemical environments with neighbors of varying composition, can give rise new emergent behaviors.

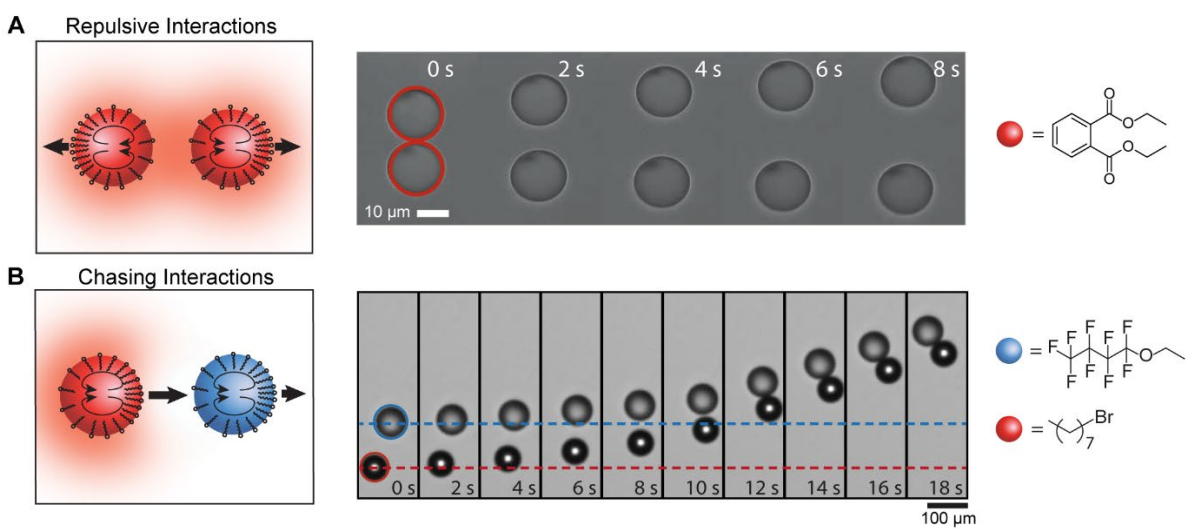

**Figure 6. Pairwise interactions between solubilizing droplets. A**, Solubilizing droplets which are repelled by their own solubilizate gradients will also repel from other droplet neighbors. As shown in the schematic, a higher concentration of solubilized oil (red shading) in between the droplets drives them to move apart. At right, optical micrographs showing diethyl phthalate droplets in SDS. The droplets were initially brough into close contact using optical tweezers. Upon release of the optical trap, the drops move away from each other. Reproduced with permission from [47], copyright American Physical Society. **B**, Schematic of droplets exhibiting chasing interactions with solubilizate concentration depicted with red shading. The blue drop (prey) uptakes solubilized oil from the red drop (predator). Both droplets are repelled from solubilized oil, setting up a chase. At right, optical micrographs show bromooctane droplets chasing ethoxynonafluorobutane droplets in 0.5 wt% Triton X-100. Reproduced with permission from [17], copyright Springer Nature Limited.

## 4.2 Many-body collective interactions

When multiple droplets are in close proximity, collective interactions can sometimes emerge that differ from two-body interactions. Krüger et al. describe how when 5CB droplets in tetradecyltrimethylammonium bromide (TTAB, 7.5 – 17.5 wt%) are spatially confined, droplets aggregate to form clusters (**Figure 7A,B**).[21] These clusters only form under specific conditions with cluster sizes and spacings being influenced by factors such as the degree of confinement through changing the height of reservoir, surfactant concentration, and buoyancy. Clusters disassemble over time as the droplets shrink through solubilization. Convective flow was shown to be a key contributor to the formation of the drop clusters. Hokmabad et al. considered this collective clustering further, examining the emergence of rotational motion due to spontaneous symmetry breaking by nonlinear hydrodynamics in the system (**Figure 7C,D**).[22] Clustering of droplets with different compositions, such "source" and "sink" drops that would also be able to chase, has also been observed (**Figure 7E**).[17] The symmetry of the source and sink droplets in the cluster impacts whether the cluster translates, rotates, or is stationary.[17] When in sufficiently high number density, active Janus droplets have also been observed to spontaneously organize into clusters of 3, 4, 5 or more droplets that rotationally spin at over 20 rotations per minute with the rotation direction determined by the Janus droplet orientation in the cluster (**Figure 7F**).[23]

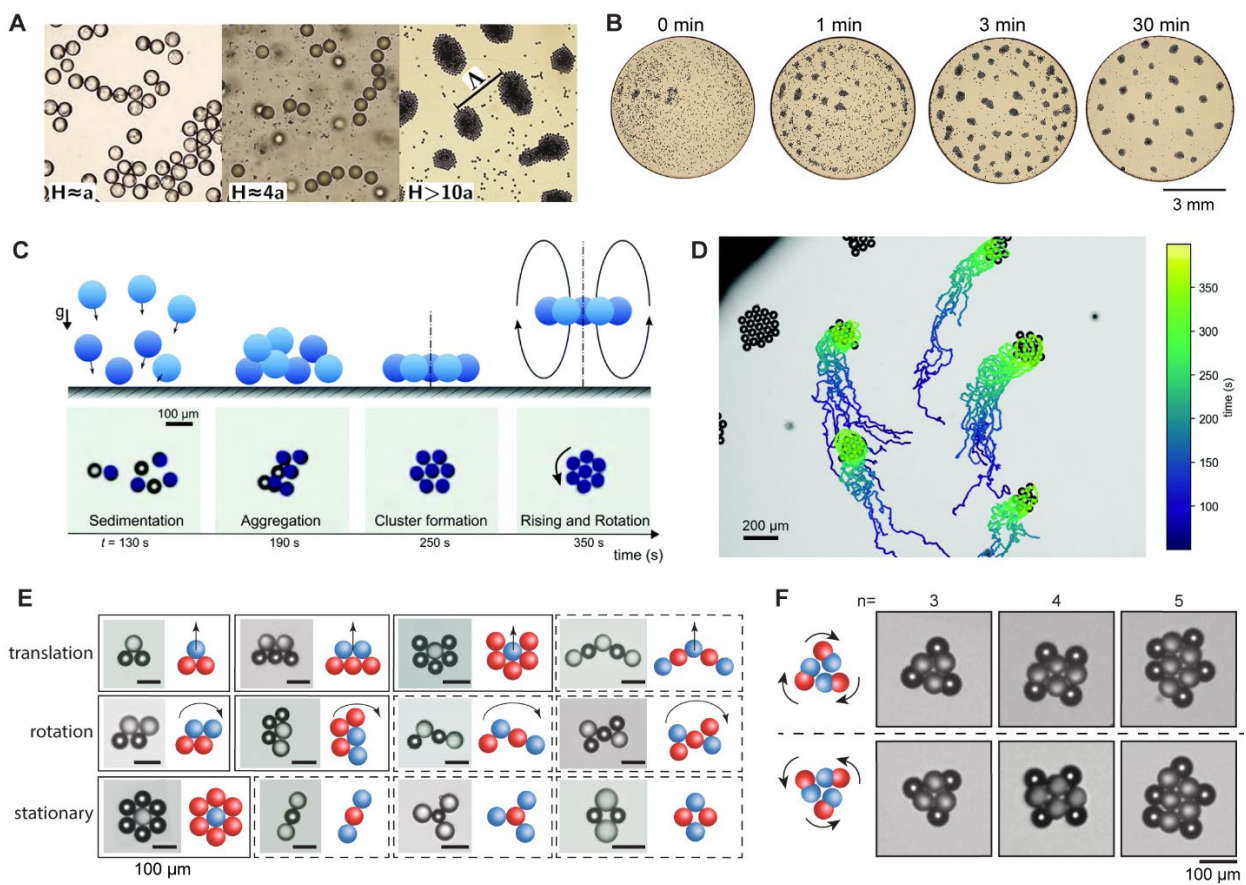

**Figure 7. Formation of droplet clusters and multibody behaviors. A**, Collective behavior of droplets can be controlled through vertical confinement where *H* represents the height of the reservoir and *a* represents diameter of droplet. Droplets shown here are 45 µm diameter for all images and are made of 5CB stabilized with TTAB. Reproduced with permission from [21] under the Creative Commons CC BY license. **B**, Time lapse showing aggregation of droplets with size and composition as described in **A** with reservoir height of 2 mm. Reproduced with permission from [21] under the Creative Commons CC BY license. **C**, Clustering of CB15 droplets in TTAB with the process and convective flows schematically represented. Reproduced with permission from [22] under a Creative Commons Attribution 3.0 Unported License. **D**, Droplets clusters as described in **C** rotate and translate. The trails of individual droplets in each cluster are color coded as a function of time. Reproduced with permission from [22] under a Creative Commons Attribution 3.0 Unported License. **E**, Bromooctane drops (red) and ethoxynonafluorobutane drops (blue)

in Triton X-100, which chase in two-body interactions, will also form larger clusters that translate, rotate, or remain stationary based on the cluster symmetry. Reproduced with permission from [17], copyright Springer Nature. **F**, Janus droplets composed of iododecane and ethoxynonafluorobutane in Triton X-100 form clusters that spin with a handedness that is dependent on the orientation of the drops in the cluster. Reproduced with permission from [23], copyright Elsevier Inc. . **B**, Time lapse showing aggregation of droplets with size and composition as described in **A** with reservoir height of 2 mm. Reproduced with permission from [21]. **C**, Clustering of CB15 droplets in TTAB with the process and convective flows schematically represented. Reproduced with permission from [22]. **D**, Droplets clusters as described in **C** rotate and translate. The trails of individual droplets in each cluster are color coded as a function of time. Reproduced with permission from [22]. **E**, Bromooctane drops (red) and ethoxynonafluorobutane drops (blue) in Triton X-100, which chase in two-body interactions, will also form larger clusters that translate, rotate, or remain stationary based on the cluster symmetry. Reproduced with permission from [17]. **F**, Janus droplets composed of iododecane and ethoxynonafluorobutane in Triton X-100 form clusters that spin with a handedness that is dependent on the orientation of the drops in the cluster. Reproduced with permission from [23].

**4.3 Droplet interactions with solid surfaces, obstacles, walls**

The presence of surfaces, obstacles, and walls have the potential to impact droplet motion by both chemical and physical means. Walls and barriers restrict the diffusion of solubilizate, which modifies chemical concentration gradients and hence interfacial tension gradients that influence droplet motion. Solid surfaces impose confinement and create hydrodynamic effects due to a no-slip condition at the interface. The size and dimensions of the sample container within which the emulsion is being studied can also matter[21] because of the generation of convective flows. For example, droplets that are denser than the continuous phase can levitate above the substrate surface[47] and the presence of a substrate can lead to symmetry breaking and fluid pumping motion even for droplets that are not self-propelled.[52,54] Research is ongoing to better understand how these effects couple and change over time and impact the droplet active behaviors.

One way to modulate the droplet motion is to position pillars of various shapes and sizes in the droplet environment (**Figure 8**). Convex pillars immersed in the continuous phase have been shown to attract droplets and, under certain conditions, induce droplets to orbit around the pillar. Jin et al. systematically studied the effect of polydimethylsiloxane pillar size (radii 50-250 µm) on the behavior of active droplets containing 5CB and 1-bromopentane with radii of 50 µm in 7.5 wt% TTAB solution.[80,84] While larger pillars prolonged the trapped orbiting of droplets, and small diameter pillars induced droplet scattering, intermediate sizes induced droplets to orbit the pillar just once (**Figure 8A,B**). By modeling the droplet as an active Brownian particle, authors were able to also estimate various quantities such as hydrodynamic and chemo-repulsive torques, diffusion coefficients, and their dependence on wall curvature. Li et al. highlights how pillars can be utilized for an application wherein self-propelling Janus droplets deposit cargo at pillars (**Figure 8C**).[108] Authors created active droplets that started as W/O and phase separated into an WW/O Janus droplet over time. The phase separation occurred due to both continuous release of ethanol into the ambient phase and the simultaneous uptake of surfactants. At different points in the droplet lifetime, the surface chemistry and the shape of the pillar influenced the pillar-droplet interactions.

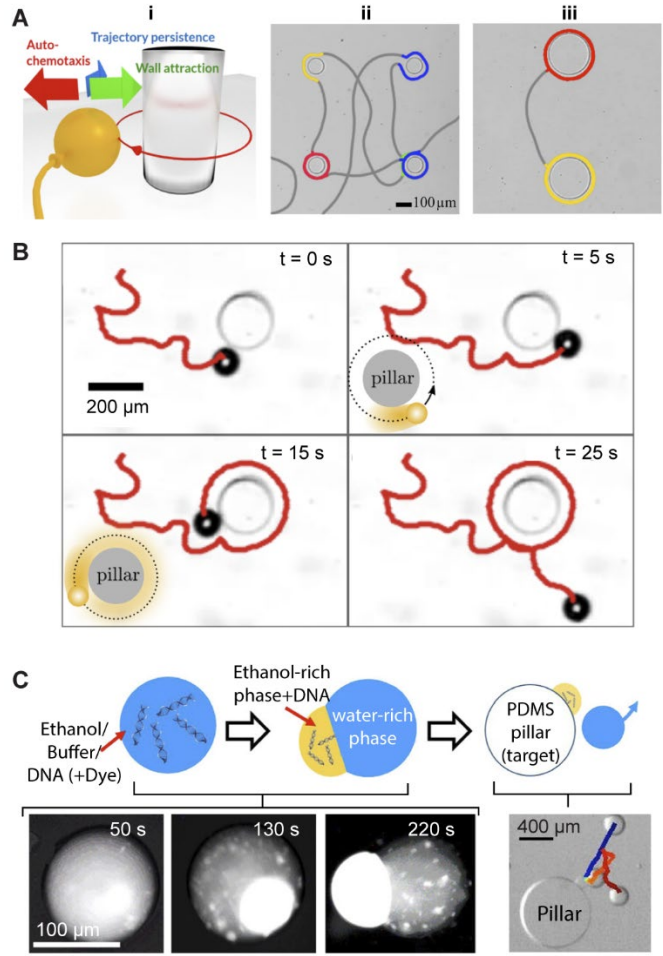

**Figure 8. Droplet interactions with pillars. A, (i)** Schematic representation of the balance of attractive and repulsive forces between the active droplet and a pillar that lead to behaviors like trapped orbiting and bouncing. **(ii, iii)** Optical micrographs of the trajectories of 5CB/BPD droplets in TTAB interacting with PDMS pillars with diameters of 100 μm and 250 μm respectively. Reproduced with permission from [84], copyright IOP Publishing. **B**, Timelapse optical micrographs and corresponding trajectories of a 5CB/BPD droplet in TTAB attaching to a PDMS pillar, orbiting the pillar, and subsequently detaching as it crosses its own repulsive trail. Reproduced with permission from [80] under the Creative Commons Attribution 4.0 International license. **C**, Top: a schematic of how an ethanol/water/DNA drop in a continuous phase of squalane/monoolein phase separates over time into a Janus droplet that self-propels and delivers the DNA-rich cargo phase onto a pillar. Below: Time-sequence fluorescence micrographs showing the drop phase separation, and a brightfield optical micrograph showing the cargo delivery onto a PDMS pillar. Images reproduced with permission from [108] under the Creative Commons CC BY license.

## 4.4 Interactions with droplet trails

Self-propelled droplets leave behind chemical trails of solubilizate or reaction products as they traverse through solution. These trails provide a short-term chemical "memory" of where droplets have been, and they can influence the motion of other droplets that contact the trail (**Figure 9**). Although the presence of chemical trails can commonly be inferred from examining the interactions of the trails with other droplets, phase contrast microscopy or fluorescent dyes (**Figure 9A**) can also be used to observe droplet trails directly.[53,62] For solubilization-driven active emulsions, the trail is composed of solubilizate, e.g., micelles that "filled" or "swollen" with the dispersed phase liquid. Since for self-propelled droplets the solubilizate is chemically repulsive (i.e. raises interfacial tension) then the droplet trail can also serve to repel other droplets that come in the vicinity (**Figure 9B**). The variation in speed of OO/W Janus droplets in Triton X-100 crossing solubilizate trails was studied as a function of trail lifetime;[23] droplets slowed down more when crossing young trails as compared to old trails, which is expected because the trail contents diffuse over time (**Figure 9C**). The timescale for trail dissipation was in good agreement with the diffusivity of the Triton X-100 surfactant micelles. Droplets were also observed to often swerve in direction when crossing trails.[17] Chemical trails also lead to self-avoiding behaviors for the propulsion of isolated droplets.[80]

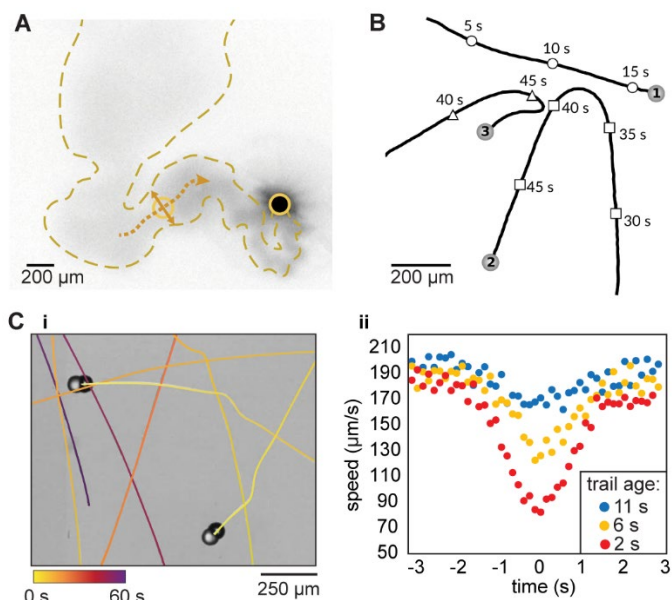

**Figure 9. Droplets interact with solubilizate trails. A**, As self-propelled droplets solubilize, they leave behind a trail of solubilizate-swollen micelles. Shown is a fluorescence micrograph of a fluorescently-dyed diethyl phthalate drop solubilizing in SDS. The width of the solubilizate trail spreads with a diffusion constant that is consistent with the size of the diethyl phthalate-swollen micelles. Reproduced from [53] under Creative Commons Attribution 4.0 International License. **B**, The trajectories of three isotropic 5CB droplets in 25 wt% TTAB are traced with the time points of observation given. Droplets are repelled by the trails of swollen micelles left behind by previous droplets passing through the same area. Reproduced with permission from [62], copyright National Academy of Sciences.[62] **C**, Janus droplets composed of iododecane and ethoxynonafluorobutane in 0.5 wt% Triton X-100 swerve and slow down when crossing other droplets' trails. **(i)** An optical micrograph of the Janus droplets with the trails color coded by lifetime according to the color scale. **(ii)** Plot of individual Janus droplets' speeds as they cross trails of different age. The time of trail crossing was defined as 0 seconds. Older trails have less impact because the oil-swollen micelles diffuse over time. Reproduced with permission from[23], copyright Elsevier Inc.

## 5 Looking Forward & Conclusions

Research on active droplets is growing rapidly, and the range of experimental conditions and material embodiments demonstrated have diversified greatly. As a deeper understanding and consensus regarding the behavior of the "simpler" single phase droplets has developed, there has been a trend towards including increasing degrees of complexity within the emulsion compositions, structural environments, chemical reactions used, and droplet morphologies. Each newly added chemical or structural variant brings additional considerations and strategies to controlling the underlying chemo-mechanical mechanisms governing droplet speeds, lifetime, distance-dependent interactions, etc. Although this review focused on experimental research, we emphasize that the field benefits greatly from collaborative interdisciplinary efforts to develop the theory and modeling of these active systems, which often involve the complex coupling between chemical gradients and hydrodynamics. Many challenges remain, especially experimental challenges. For example, it is essentially impossible to control for changes in chemical and physical variables independently in an active emulsion; for instance, by using additives to change viscosity, we may inadvertently change the interfacial tensions or solubilization kinetics. Experimental characterization of the structure and dynamics of liquid-liquid interfaces is often difficult or imprecise. Even well-established techniques like the pendant drop method used to measure "bulk" interfacial tensions may have limitations regarding our understanding of the interfacial tension and dynamics of microscale droplets.[114,115] Despite the hurdles, we remain excited about the prospects of future discoveries in the field of active droplets and the new "life" it has infused into fundamental emulsion science research.

## Abbreviations and Symbols

$\gamma$ – interfacial tension
5CB – 4-cyano-4'-pentylbiphenyl (or pentylcyanobiphenyl)
5CT – 4'-cyano-4-*n*-pentyl-*p*-terphenyl
7CB – 4-heptyl-4'-cyanobiphenyl
8CB – 4-*n*-octyl-4'-cyanobiphenyl
8OCB – 4-*n*-octyloxy-4'-cyanobiphenyl
ATPS – aqueous two-phase system
BHAB – 4-butyl-4'-hydroxyazobenzene
BMAB – 4-butyl-4'-methoxyazobenzene
BPD – bromopentadecane
BZ – Belousov-Zhabotinsky (reaction)
C12-HPTS - 8-hydroxypyren-1,3,6-trisulfonic acid (derivatized with dodecyl groups)
CB15 – (*S*)-4-cyano-4'-(2-methylbutyl)biphenyl
CMC – critical micelle concentration
CTAB – cetyltrimethylammonium bromide
DCE – dichloroethane
DEHPA – di-(2-ethylhexyl)phosphoric acid
DEP – diethyl phthalate
DSGC – disodium glycate
E7 – Merck trade name for mixture of 51% 5CB, 25% 7CB, 16% 8OCB, 8% 5CT
EFB - ethoxynonafluorobutane
HBA – heptyloxybenzaldehyde
HDA – 2-hexyldecanoic acid
HEPES - 4-(2-hydroxyethyl)-1-piperazineethanesulfonic acid
HFE-7500 – 3-ethoxy-1,1,1,2,3,4,4,5,5,6,6,6-dodecafluoro-2-(trifluoromethyl)-hexane
HLB – hydrophilic lipophilic balance
LC – liquid crystal
LCLC – lyotropic chromonic liquid crystal
LOETAB - Lauroyloxyethylene-*N,N,N*-trimethylammonium Bromide
M280 – 2.8-μm superparamagnetic beads (Bangs Lab)
MBBA – *N*-(4-methoxybenzylidene)-4-butylaniline

O/W – oil-in-water
PBS – phosphate buffered saline
PDMS – polydimethylsiloxane
PEG – poly(ethylene glycol)
PFPE – perfluoropolyether
PVA – poly(vinyl alcohol)
R811 – (*R*)-2-octyl 4-[4-(hexyloxy)benzoyloxy]benzoate
SDS – sodium dodecyl sulfate
TB – terrific broth
TTAB – trimethyltetradecylammonium bromide
W/O – water-in-oil
W/O/W – water-in-oil-in-water

**Declaration of competing interests.** The authors declare no competing interests.

**Acknowledgements.** This work was supported by the Army Research Office (W911NF-18-1-0414) and the Marion Milligan Mason Award, AAAS (236764).

**Graphical abstract**

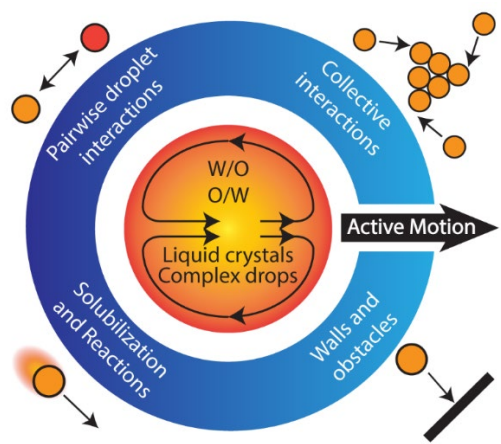

**References selected for annotation**

Maass, C. C.; Krüger, C.; Herminghaus, S.; Bahr, C. Swimming Droplets. *Annu. Rev. Condens. Matter Phys.* **2016**, *7* (1), 171–193.
- Review paper describing the progress of active droplets research up to 2016

Hanczyc, M. M.; Toyota, T.; Ikegami, T.; Packard, N.; Sugawara, T. Fatty Acid Chemistry At The Oil-Water Interface: Self-Propelled Oil Droplets. *J. Am. Chem. Soc.* **2007**, *129* (30), 9386–9391.
- Report describing active droplets propelled by interfacial reaction mechanism

Izri, Z.; Van Der Linden, M. N.; Michelin, S.; Dauchot, O. Self-Propulsion of Pure Water Droplets by Spontaneous Marangoni-Stress-Driven Motion. *Phys. Rev. Lett.* **2014**, *113* (24).
- Seminal paper describing isotropic water droplets that self-propel by solubilization